\documentclass[aps,prl,twocolumn]{revtex4-1}
\usepackage{mathrsfs}
\usepackage{bbm}
\usepackage{amsmath}
\usepackage{amssymb}
\usepackage{graphicx}
\usepackage{MnSymbol}
\pdfoutput=1

\newcommand{\mca}{\mathcal}

\newcommand{\mfr}{\mathfrak}

\makeatletter
\usepackage{hyperref}
\usepackage{color}
\definecolor{myblue}{RGB}{0,50,200}
\hypersetup{
	colorlinks,
	citecolor = myblue,
	linkcolor = myblue,
	urlcolor = myblue
}

\newcommand{\var}[1]{\langle\langle {#1} \rangle\rangle}
\newcommand{\mean}[1]{\langle {#1} \rangle}
\newcommand{\braket}[1]{\left( {#1} \right)}
\newcommand{\abs}[1]{\left| {#1} \right|}

\begin{document}
\title{Unified Approach to Classical Speed Limit and Thermodynamic Uncertainty Relation}
\author{Van Tuan Vo}
\email{tuan@biom.t.u-tokyo.ac.jp}
\affiliation{Department of Information and Communication Engineering, Graduate School of Information Science and Technology, The University of Tokyo, Tokyo 113-8656, Japan}

\author{Tan Van Vu}
\email{tan@biom.t.u-tokyo.ac.jp}
\affiliation{Department of Information and Communication Engineering, Graduate School of Information Science and Technology, The University of Tokyo, Tokyo 113-8656, Japan}

\author{Yoshihiko Hasegawa}
\email{hasegawa@biom.t.u-tokyo.ac.jp}
\affiliation{Department of Information and Communication Engineering, Graduate School of Information Science and Technology, The University of Tokyo, Tokyo 113-8656, Japan}

\begin{abstract}
The total entropy production quantifies the extent of irreversibility in thermodynamic systems, which is nonnegative for any feasible dynamics. 
When additional information such as the initial and final states or moments of an observable is available, it is known that tighter lower bounds on the entropy production exist according to the classical speed limits and the thermodynamic uncertainty relations. 
Here, we obtain a universal lower bound on the total entropy production in terms of probability distributions of an observable in the time forward and backward processes.
For a particular case, we show that our universal relation reduces to a classical speed limit, imposing a constraint on the speed of the system's evolution in terms of the Hatano--Sasa entropy production. Notably, the newly obtained classical speed limit is tighter than the previously reported bound by a constant factor. 
Moreover, we demonstrate that a generalized thermodynamic uncertainty relation can be derived from another particular case of the universal relation. 
Our new uncertainty relation holds for systems with time-reversal symmetry breaking and recovers several existing bounds. 
Our approach provides a unified perspective on two closely related classes of inequality: classical speed limits and thermodynamic uncertainty relations. 
\end{abstract}

\maketitle

\emph{Introduction.}---Irreversibility is one of the cornerstones for understanding the physical mechanisms of nonequilibrium systems and closely connected to the energy dissipation.
Typically, the irreversibility is measured by the relative entropy between the probability distributions of system trajectories in the forward and the time-reversed processes \cite{Seifert:2012:RPP}.
If the dynamics obey the local detailed balance condition, this relative entropy is thermodynamically related to the total entropy production.
According to the second law of thermodynamics, the total entropy production is expected to be nonnegative.
Although this zero lower bound is universally valid, it is irrelevant as to how far from equilibrium a given system is.

Recently, one of the most sought-after challenges is to tighten the constraints on the total entropy production.
Inspired by research pertaining to the quantum regime, classical speed limits (CSLs) have been derived for Markov jump processes \cite{Shiraishi:2018:PRL}, Fokker--Planck dynamics \cite{Aurell:2011:PRL, Aurell:2012:JSP, Dechant:2019:arxiv, Ito:2020:PRX}, and Liouville dynamics \cite{Okuyama:2018:PRL, Shanahan:2018:PRL}.
For Markov jump processes, the CSL implies that given a distance between the initial and final distributions, the faster the speed of evolution, the more entropy production is required.
The CSL can be interpreted as a lower bound of the total entropy production in terms of the state transformation speed.
Another class of constraints, which is known as thermodynamic uncertainty relations (TURs), has been discovered for both classical and quantum regimes \cite{Barato:2015:PRL, Gingrich:2016:PRL, Pietzonka:2016:PRE, Horowitz:2017:PRE, Dechant:2020:PNASUSA, Hasegawa:2019:PRE, Hasegawa:2019:PRL, VanVu:2019:PRE, VanVu:2020:PRR, Hasegawa:2020:QTUR} (see \cite{Horowitz:2020:NP} for a review). 
For time-reversal symmetric systems, TURs impose a bound on the average entropy production in terms of current precision.
TURs have a wide range of practical applications, particularly for inferring of dissipation without requiring the detailed knowledge of the underlying dynamics of the system \cite{Li:2019:NC, Manikandan:2020:PRL, VanVu:2020:PRE, Otsubo:2020:PRE}.
In light of these results, the question arises whether there exists a universal relation that reveals the origin of the trade-offs among irreversibility, precision, and the speed at which a system evolves.

In this letter, we answer this question by deriving a universal relation between the irreversibility of a system and its physical observable.
Specifically, we prove that the degree of irreversibility of the system is lower bounded by a quantity involving the probability distributions of the observable in the forward and backward processes.
Recognizing the similarity between CSL and TUR in the sense of constraints on the irreversibility, we show that these trade-offs can be derived from the universal relation.
First, by quantifying the irreversibility using the Hatano--Sasa entropy production, we obtain a CSL for continuous-time Markov jump processes.
The operational time is bounded from below by a combination of the Hatano--Sasa entropy production, the dynamical activity, and the distance between the initial and final states.
Notably, the obtained bound is tighter by a specific constant factor than that reported in Ref.~\cite{Shiraishi:2018:PRL}.
Second, as another corollary of the universal relation, we obtain a generalized TUR for systems in which the time-reversal symmetry is broken.
The degree of irreversibility is constrained from below by the mean and variance of the observable in both forward and backward processes, reflecting the presence of the time-reversal symmetry breaking.
Our TUR can be considered as a generalization of TURs, since it recovers several TURs for the time-reversal symmetric systems considered in Refs.~\cite{Hasegawa:2019:PRL, Timpanaro:2019:PRL}.
From the unified perspective on CSL and TURs, we hope that this study will provide a promising avenue for deriving more types of trade-offs.
 
\emph{Lower bound on the irreversibility.}---We start by considering a stochastic system evolving in the phase space during the time interval of $[0,\tau]$.
Let $\omega$ be the trajectory and $P_{\rm F}( \omega)$ be the probability of observing $\omega$ in a time-forward process. 
The conjugate of this forward process is a backward process with the time-reversed trajectory $\omega^\dagger$.
Here, the superscript $\dagger$ denotes the time-reversal operator, which reverses the order of the states in time, and changes their sign according to their parities (such as their velocities or in the presence of a magnetic field). 
Let $P_{\rm B}( \omega^\dagger)$ denote the probability that the system could take the time-reversed trajectory $\omega^\dagger$ during the backward process.
Hereafter, we use the subscripts ${\rm F}$ and ${\rm B}$ to refer to the forward and backward processes, respectively.

We define a trajectory-dependent quantity, $\sigma$, which encodes the irreversibility of the trajectory $\omega$ as
\begin{align} \label{Entropy1}
    \sigma( \omega) = \ln \frac{P_{\rm F}( \omega)}{P_{\rm B}( \omega^\dagger)}.
\end{align}
Depending on the setup of the backward process, we have different physical interpretations of $\sigma$ \cite{Seifert:2012:RPP}. 
If the backward process is the time reversal of the forward process and the dynamics obey the local detailed balance condition, $\sigma$ is equivalent to the total entropy production \cite{Seifert:2005:PRL}.
When the dynamics of the backward process are the dual ones \cite{Esposito:2010:PRL}, $\sigma$ corresponds to the nonadiabatic contribution of the total entropy production.
In the presence of measurements and feedback, $\sigma$ includes the total entropy production and an information term \cite{Sagawa:2012:PRE, Potts:2018:PRL}.

The quantity of interest is the average of $\sigma$ over the ensemble of trajectories, which can be expressed as 
\begin{align}\label{entropy}
    \mean{\sigma}_{\rm F} &=  \sum_ \omega P_{\rm F}( \omega)\ln\frac{P_{\rm F}( \omega)}{P_{\rm B}( \omega^\dagger)} 
    \equiv  D[P_{\rm F}( \omega)||P_{\rm B}( \omega^\dagger)],
\end{align}
where $D$ denotes the relative entropy, and $\mean{\cdot}$ denotes the ensemble average.
In general, it is infeasible to fully determine the probability of all trajectories, due to the presence of hidden degrees of freedom and the existence of exponentially rare events.
Moreover, it is impractical to directly calculate the relative entropy due to heavy computational costs of processing high-dimensional data. 
In what follows, we derive an accessible lower bound on the ensemble average of $\sigma$.

Let us consider an observable $\phi(\omega)$, which is a trajectory-dependent quantity.
The observable has the probability distribution $P(\phi) = \sum_\omega P(\omega) \delta(\phi, \phi(\omega))$, where $\delta (x, y)$ is the Kronecker delta ($\delta(x,y)=1$ for $x=y$, and $0$ otherwise).
We assume that the observable in the backward process is the same as the forward one, i.e., $\phi^\dagger (\omega^\dagger)= \phi(\omega^\dagger)$, and $\phi^\dagger$ is uniquely determined if $\phi$ is given.
Following the chain rule for the divergence of a joint probability distribution \cite{Cover:2006:book}, we obtain
\begin{align}\label{Chain_Rule}
	&D[P_{\rm F}(\omega)||P_{\rm B}(\omega^\dagger)]  +  D[P_{\rm F}( \phi| \omega)||P_{\rm B}( \phi^\dagger|\omega^\dagger)] \nonumber \\
	&= D[P_{\rm F}( \phi) || P_{\rm B}( \phi^\dagger)] +  D[P_{\rm F}(\omega|\phi) || P_{\rm B}( \omega^\dagger|\phi^\dagger)], 
\end{align}
where $D[P(y|x) || Q(y|x)] \equiv \sum_x P(x) \sum_y P(y|x) \ln \frac{P(y|x)}{Q(y|x)}$ is the conditional divergence between the joint probability distributions $P(x,y)$ and $Q(x,y)$. Due to the nonnegativity of the conditional divergences and $P_{\rm F}( \phi| \omega)= P_{\rm B}( \phi^\dagger|\omega^\dagger)= \delta(\phi, \phi(\omega))$, Eq.~\eqref{Chain_Rule} yields the following information-processing inequality \cite{Cover:2006:book}:
\begin{align} \label{Information_Processing}
    D[P_{\rm F}(\omega)||P_{\rm B}(\omega^\dagger)] \geq  D[P_{\rm F}( \phi) || P_{\rm B}(\phi^\dagger)].
\end{align}
Hence, the most information about the irreversibility that one can extract from the observable is the relative entropy $D[P_{\rm F}( \phi) || P_{\rm B}( \phi^\dagger)]$.
When $\phi(\omega)$ is a reduced trajectory, $D[P_{\rm F}( \phi) || P_{\rm B}( \phi^\dagger)]$ is equivalent to the coarse-grained relative entropy \cite{Kawai:2007:PRL, Roldan:2010:PRL, Roldan:2012:PRE}.
By noting that $p\ln \frac{p}{q} -p + q = \int_0^1 \frac{\theta (p - q)^2}{(1- \theta)p+ \theta q} d \theta$, we find
\begin{align} 
	 D[P_{\rm F}( \phi) || P_{\rm B}( \phi^\dagger)]
	= \int_0^1 \sum_ \phi \frac{\theta(P_{\rm F}( \phi) - P_{\rm B}( \phi^\dagger))^2}{P_\theta(\phi)} d\theta, \nonumber
\end{align}
where $P_\theta(\phi) = (1- \theta) P_{\rm F}( \phi) + \theta P_{\rm B}( \phi^\dagger)$. 
We apply the Cauchy--Schwarz inequality to arrive at
\begin{align}
\mean{\sigma}_{\rm F} \geq \int_0^1  \frac{ \theta \braket{\sum_ \phi\abs{f_\theta(\phi) (P_{\rm F}( \phi) - P_{\rm B}( \phi^\dagger))}}^2}{\sum_\phi f_\theta(\phi)^2 P_\theta(\phi) } d\theta,\label{First_Bound}
\end{align}
where $f_\theta(\phi)$ is any function of $\phi$ and $\theta$ that satisfies $\sum_\phi f_\theta(\phi)^2 P_\theta(\phi) \neq 0$.
This relation connects the degree of reversibility and statistical values of the observable, illustrating the fact that details of the process are incompletely represented by the observable.
This result holds not only for classical stochastic dynamics but also for deterministic systems, non-Markovian processes, and quantum trajectories.
Since the term on the right-hand side of Eq.~\eqref{First_Bound} is nonnegative, this relation tells us more about the degree of irreversibility than the second law of thermodynamics does.
We will show that, by choosing an appropriate function $f_\theta(\phi)$, one can obtain a tighter bound for the CSL and a generalized TUR.

\emph{Classical speed limit.}---We consider a continuous-time Markov process with discrete states $\{1, \dots, N \}$.
The time evolution of the probability $p_n(t)$ to find the system in the state $n$ at the time $t$ is described by the following master equation:
\begin{align}
	 \dot{p}_n(t) = \sum_{m} R_{nm}(t) p_m(t),
\end{align}
where the transition from the state $m$ to the state $n$ occurs at rate $R_{nm}(t)$.
The transition rates satisfy the normalization condition, $\sum_{n} R_{nm} = 0$, to ensure the conservation of the total probability.
We assume that the transition rate from the state $n$ to state $m$ $(\neq n)$ is nonnegative. The total entropy production rate is given by \cite{Seifert:2012:RPP}
\begin{align}
	\dot{\Sigma}(t) &\equiv \ \sum_{n, m} R_{m n}(t) p_{n}(t) \ln \frac{R_{mn(t)} p_n(t) }{R_{nm(t)}p_m(t)}.
\end{align}

We assume that a unique instantaneous stationary distribution $p_n^{\textrm{ss}}(t)$ exists, which satisfies $\sum_m R_{nm} p_m^{\textrm{ss}} = 0$.
The total entropy production rate can be decomposed as \cite{Hatano:2001:PRL, Esposito:2010:PRL}
\begin{align}
    \dot{\Sigma}(t) = \dot{\Sigma}^{\rm{A}}(t) + \dot{\Sigma}^{\rm{HS}}(t),
\end{align}
where $\dot{\Sigma}^{\rm{A}}(t)$ denotes the adiabatic entropy production rate given by
\begin{align}
     \dot{\Sigma}^{\mathrm{A}}(t) &\equiv  \sum_{n, m} R_{m n}(t) p_{n}(t) \ln \frac{R_{mn}(t)p_n^\textrm{ss}(t)}{R_{nm}(t)p_m^{\textrm{ss}}(t)}
\end{align}
and $\dot{\Sigma}^{\rm{HS}}(t)$ is the nonadiabatic contribution, also known as the Hatano--Sasa entropy production rate:
\begin{align}
    \dot{\Sigma}^{\mathrm{HS}}(t) &\equiv \sum_{n, m} R_{m n}(t) p_{n}(t) \ln \frac{p_m^\textrm{ss}(t) p_n(t)}{p_n^{\textrm{ss}}(t)p_m(t)}.
\end{align}
The Hatano--Sasa entropy production rate is less than or equal to the total entropy production rate, due to the nonnegativity of the adiabatic contribution.
The equality is attained if the transition rates satisfy the detailed balance condition.

Now, let us consider a backward process with the dual dynamics \cite{Seifert:2012:RPP}, whose transition rates are defined as
\begin{align}
	\tilde{R}_{nm}\equiv\frac{R_{m n} p_{n}^{\mathrm{ss}}}{p_{m}^{\mathrm{ss}}}.
\end{align} 
Obviously, the escape rates of the original and dual dynamics are equal, $\tilde{R}_{n n} = R_{n n} $.
Since the transition rates of the dual dynamics also satisfy the normalization condition, $\sum_{m}\tilde{R}_{mn} = 0$, we obtain
\begin{align}
	\sum_{m(\neq n)}\tilde{R}_{mn} = - \tilde{R}_{n n} = - R_{n n} =  \sum_{m(\neq n)} R_{mn}.
\end{align}
Let $\mathcal A(t) \equiv \sum_{n \neq m } R_{mn}(t) p_n(t)$ denote the dynamical activity, which describes the frequency of jumps \cite{Baiesi:2009:PRL, Baiesi:2009:JSP}. 
The dynamical activities in the dual dynamics and the original ones are equal:
\begin{align}
	 \tilde{ \mathcal A}\equiv \sum_{ n\neq m} \tilde{R}_{m n} p_n = \sum_{ n\neq m} R_{m n} p_n = \mathcal A.
\end{align}

Indeed, dual dynamics provide another way to express the Hatano--Sasa entropy production rate:
\begin{align}
	\dot{\Sigma}^{\mathrm{HS}}=\sum_{n \neq m} R_{m n} p_{n} \ln \frac{R_{m n} p_{n}}{\tilde{R}_{nm} p_{m}}.
\end{align}
To be consistent with the previous notation, we denote $\omega_{nm}$ as the forward transition from the state $n$ to state $m$ ($n \neq m$) and introduce a probability distribution $P_{\rm F}(\omega_{nm}) \equiv \mathcal A ^{-1}R_{mn}p_n$. 
The probability distribution of the backward transition $\omega_{nm}^\dagger$ is then defined as $P_{\rm B}(\omega_{nm}^\dagger) \equiv \mathcal A^{-1} \tilde{R}_{nm} p_{m}$.
Then, we can rewrite the Hatano--Sasa entropy production rate as
\begin{align}
	\dot{\Sigma}^{\mathrm{HS}}= \mathcal A D[P_{\rm F}(\omega_{mn})||P_{\rm B}(\omega_{mn}^\dagger)].
\end{align}
If we set the function $f_\theta(\phi) = 1 $ in Eq.~\eqref{First_Bound}, we find that  $D[P_{\rm F}(\omega_{mn})||P_{\rm B}(\omega_{mn}^\dagger)]$ is lower bounded by
\begin{align} \label{Cauchy_Schwarz_CSL}
\geq \int_0^1 \frac{ \theta \left( \sum_ \phi |P_{\rm F}( \phi) - P_{\rm B}( \phi^\dagger) |\right)^2}{ \sum_ \phi P_\theta(\phi)} d \theta.
\end{align}
Providing that $\sum_ \phi P_\theta(\phi) = \sum_ \phi [(1- \theta) P_{\rm F}( \phi) + \theta P_{\rm B}( \phi^\dagger)] = 1$, we obtain
\begin{align}
		\dot{\Sigma}^{\mathrm{HS}} \geq \frac{\mathcal A}{2}\left( \sum_ \phi |P_{\rm F}( \phi) - P_{\rm B}( \phi^\dagger) |\right)^2,
\end{align}
which is a lower bound on the Hatano--Sasa entropy production rate in terms of the dynamical activity and the total variation distance between the distributions of the observable.
We choose the observable $\phi(\omega_{mn})$ to be the trajectory itself, i.e., $\phi(\omega_{mn})=\omega_{mn}$.
In that case, the total variation distance becomes
 \begin{align}
	&\sum_\phi | P_{\rm F}(\phi) - P_{\rm B}(\phi^\dagger)| = \frac{1}{\mathcal A} \sum_{n \neq m} | R_{m n} p_{n} - \tilde{R}_{nm} p_{m}|.
\end{align}
Applying the triangle inequality, we get
\begin{align}
	&\sqrt{2 \mathcal A\dot{\Sigma}^{\mathrm{HS}} }  \geq \sum_{n \neq m} | R_{m n} p_{n} - \tilde{R}_{nm} p_{m}| \nonumber\\
	&\geq \sum_n \abs{\sum_{m (\neq n)} R_{nm} p_{m}  - \tilde{R}_{mn} p_{n} } =  \sum_n \abs{ \dot p_n},
\end{align}
Integrating over time from $t=0$ to $t = \tau$, we find
\begin{align}
	&\sum_n \abs{p_n(0) - p_n(\tau)} \leq \sum_n \int_0^\tau \abs{\dot  p_{n}(t)} dt \nonumber \\
	&\leq \int_0^\tau \sqrt{2 \mathcal A(t)\dot{\Sigma}^{\mathrm{HS}}(t)} dt \leq \sqrt{2 \tau \mean{\mathcal A}_\tau \Sigma ^{\mathrm{HS}}} \label{Pre_CSL},
\end{align}
where $\Sigma^{\mathrm{HS}} \equiv \int_{0}^{\tau} \dot{\Sigma}(t) dt$ is the Hatano--Sasa entropy production, and $\langle \mathcal A\rangle_{\tau} \equiv \tau^{-1} \int_{0}^{\tau}  \mathcal A(t)dt$ is the time average of the dynamical activity.
We can rewrite Eq.~\eqref{Pre_CSL} to obtain the minimal time required for the system to evolve from the initial to the final configuration as
\begin{align}{\label{CSL}}
\hat{\tau} \equiv \frac{\mca{L}(\boldsymbol p(0), \boldsymbol p(\tau))^2}{2 \Sigma^{\rm HS} \mean{\mathcal A}_\tau} \leq \tau,
\end{align}
where $\mca{L}\left(\boldsymbol{p}, \boldsymbol{q}\right):=\sum_{n}\left|p_{n}-q_{n}\right|$ is the total variation distance between distributions $\boldsymbol{p}$ and $\boldsymbol{q}$.
In Eq.~\eqref{CSL}, the dynamical activity contributes as a timescale of the state transformation, which is similar to role of the Planck constant in the quantum speed limit \cite{Shiraishi:2018:PRL}.
The bound in \eqref{CSL} does indeed constrain the Hatano--Sasa entropy production by the speed of evolution, even when the detailed balance condition is not fulfilled.
We stress that this speed limit is tighter than the relation found in Ref.~\cite{Shiraishi:2018:PRL}, where the transformation time was bounded by $c^{*}\hat{\tau}$, where $c^* \approx 0.896$.
The stronger inequality gives us the edge over many applications, such as for entropy production estimations, model validations, and design criteria for mesoscopic devices.

\emph{Thermodynamic uncertainty relation.}---Inspired by Ref.~\cite{Nishiyama:2020:E}, we derive a new TUR for systems with time symmetry breaking.
Let us consider an observable $\phi(\omega)$ satisfying $\phi(\omega) = \epsilon \phi(\omega^\dagger)$, where the parity operator is $\epsilon = 1$ $ (-1)$ for an even (odd) observable under time-reversal.
We note that the observable does not have to be a current. 
We choose $f_\theta(\phi) = \phi - \psi_\theta$ with $\psi_\theta \equiv (1- \theta) \mean{\phi}_{\rm F} + \theta \epsilon \mean{\phi}_{\rm B}$, which can be substituted into Eq.~\eqref{First_Bound} to yield
\begin{align} \label{pre_TUR1}
 \mean{\sigma}_{\rm F}\geq  \int_0^1 \frac{\theta \braket{\sum_ \phi \abs{(\phi - \psi_\theta)(P_{\rm F}( \phi) - P_{\rm B}( \phi^\dagger) )} }^2}{ \sum_ \phi (\phi - \psi_\theta)^2 P_\theta(\phi) } d\theta.
 \end{align}
By recognizing that
\begin{align} \label{pre_TUR2}
	\sum_\phi \abs{(\phi -\psi_\theta )(P_{\rm F}(\phi) - P_{\rm B}(\phi^\dagger))}
	\geq \abs{\mean{\phi}_{\rm F} - \epsilon\mean{\phi}_{\rm B}},
\end{align}
we obtain the following TUR:
\begin{align}\label{TUR}
	\mean{\sigma}_{\rm F} \geq  \int_0^1 \frac{\theta (\mean{\phi}_{\rm F} - \epsilon\mean{\phi}_{\rm B})^2 }{\var{\phi}_\theta} d\theta,
\end{align}
with $\var{\phi}_\theta = (1-\theta) \var{\phi}_{\rm F} + \theta \var{\phi}_{\rm B}  + \theta(1-\theta)(\mean{\phi}_{\rm F} - \epsilon\mean{\phi}_{\rm B})^2$, where $\var{\phi}$ denotes the variance of the observable.
We note that the right-hand side of Eq.~\eqref{TUR} contains only the average and variance of the observable.
The explicit form of the lower bound is $  \frac{1}{2} \frac{\mfr{a} - 2 \var{\phi}_{\rm F}}{\mfr{b}} \ln{\frac{\mfr{a}+ \mfr{b}}{\mfr{a} - \mfr{b}}} +  \frac{1}{2}\ln{\frac{\var{\phi}_{\rm F}}{\var{\phi}_{\rm B}}}$ with $\mfr{a} \equiv (\mean{\phi}_{\rm F} - \epsilon\mean{\phi}_{\rm B})^2 + \var{\phi}_{\rm F} +\var{\phi}_{\rm B}$ and $\mfr{b} \equiv \sqrt{\mfr{a}^2 - 4\var{\phi}_{\rm F}\var{\phi}_{\rm B}}$, which is a particular case of the lower bound on the relative entropy derived in Ref.~\cite{Nishiyama:2020:E}.

We remark that our result is valid for a wide variety of systems with broken time-reversal symmetry, including magnetic fields, time-antisymmetric external protocols, feedback, and underdamped Langevin dynamics.
This framework could be extended to other phenomena containing symmetry breaking, such as the breaking of space inversion by external factors in equilibrium states \cite{Lacoste:2014:PRL}.
In the general settings, we find that, unlike conventional TURs, the observable in the backward process is indispensable.
This perspective is consistent with those reported in Refs.~\cite{Proesmans:2019:JSM, Potts:2019:PRE, Falasco:2020:NJP}.
However, the previous TURs required summing of the entropy production of the forward and backward processes $\mean{\sigma}_{\rm F} + \mean{\sigma}_{\rm B}$, making it infeasible to infer the entropy production $\mean{\sigma}_{\rm F}$ alone.
From our result, we note that one can obtain a tighter bound on the summed of entropy productions than those found in Refs.~\cite{Proesmans:2019:JSM, Potts:2019:PRE, Falasco:2020:NJP}.
It is also worth emphasizing that the observable could be symmetric under time reversal, which cannot be handled by the previous approaches.
There was another attempt to modify the TUR for systems with time-dependent driving, which considered the response of the observable to a small change of speed and time scale in \cite{Koyuk:2020:arxiv}.
Nonetheless, this method was restricted to overdamped Langevin dynamics and required a precise control of the system as well as measurements.

Now, we will show that some conventional TURs can be recovered via our the derived TUR. 
Let us consider the special case where the control protocol is time-symmetric, $P_{\rm F}(\omega) = P_{\rm B} (\omega)$, and the observable $\phi$ is odd under time-reversal, i.e., $\phi(\omega) = -\phi(\omega^\dagger)$. In this case, we have $\mean{\phi}_{\rm F} = \mean{\phi}_{\rm B}$ and $\var{\phi}_{\rm F} = \var{\phi}_{\rm B}$.
Then, our TUR reduces to $\mean{\sigma}_{\rm F} \geq \frac{1}{g(\phi)} \ln \frac{g(\phi) + 1}{g(\phi) -1 }$, where $g(\phi) = \sqrt{1 + \frac{\var{\phi}_{\rm F}}{\mean{\phi}_{\rm F}^2}}$.
This can be rewritten as
\begin{align}\label{CSCH}
	\frac{\var{\phi}_{\rm F}}{\mean{\phi}_{\rm F}^2} \geq \textrm{csch}^2 \left[h\left(\frac{\mean{\sigma}_{\rm F}}{2} \right)\right] \geq \frac{2}{e ^{\mean{\sigma}_{\rm F}} - 1},
\end{align}
where $\textrm{csch}(x)$ is a hyperbolic cosecant, and $h(x)$ is the inverse of the function $x \tanh (x)$. Equation \eqref{CSCH} is exactly the tightest lower bound introduced in Refs. \cite{Timpanaro:2019:PRL, Hasegawa:2019:PRL}.

Let us examine the conditions for equality in our TUR.
From Eq.~\eqref{First_Bound} and \eqref{pre_TUR2}, the equality will only be attained if the following two conditions are met.
The first condition is $P_{\rm F}(\omega|\phi) = P_{\rm B}( \omega^\dagger|\phi^\dagger)$ for all $\omega$ and $\phi$.
The second is the existence of a function $k(\theta)$ such that $k(\theta) = \frac{P_{\rm F}(\phi) - P_{\rm B}(\phi^\dagger)}{(\phi - \psi_\theta)P_\theta(\phi)}$ for all $\phi$ and $\theta \in [0,1]$. 
We remark that the above equality constraints may be satisfied regardless of the magnitude of $\mean{\sigma}_{\rm F}$, which offers a significant advantage in terms of estimating the irreversibility of stochastic processes. In Ref. \cite{Supp.PhysRev}, we illustrate that equality conditions are satisfied by the generalized Szilard engine with measurement errors \cite{Sagawa:2012:PRE}, in which we choose the work as the observable.

\emph{Conclusions and discussion.}---In this letter, we identified the physical essence of a relation between the irreversibility and a physical observable.
From the derived universal relation \eqref{First_Bound}, we obtained two new bounds on the degree of irreversibility in the forms of the CSL and TUR, which illustrate the nature that large dissipation is required to achieve high speed and exquisite precision.
As a future study, it would be interesting to apply these relations to derive the physical limits of the heat engine performance \cite{Pietzonka:2018:PRL} and irreversible computations \cite{Berut:2012:N, Gavrilov:2017:PNASUSA}.

We anticipate that the work presented here will shed light on new ways to derive rigorous relations among many physical quantities.
Since our approach deploys the monotonicity of the relative entropy under information processing, its application is not restricted to systems that contain the notion of time evolution.
This perspective suggests that our method can be used to derive relations between relevant characteristics of the broken symmetry in the configurations of both classical and quantum systems, such as magnetic systems and nematic liquid crystals \cite{Hurtado:2011:PNASUSA, Lacoste:2014:PRL}.
Such relations are important not only in physics but also in the machine-learning field, in which the evaluation of the relative entropy between the probability distributions of the data and model is a major problem \cite{Bishop:2006:book, Goodfellow:2016:book, Nguyen:2010:ITIT}.

\emph{Acknowledgments.}---This work was supported by the Ministry of Education, Culture, Sports, Science and Technology (MEXT) KAKENHI Grant No.~JP19K12153.

\end{document}


\title{Supplementary Material for\\ 
``Unified Approach to Classical Speed Limit and Thermodynamic Uncertainty Relation''}
\author{Van Tuan Vo}
\email{tuan@biom.t.u-tokyo.ac.jp}
\affiliation{Department of Information and Communication Engineering, Graduate School of Information Science and Technology, The University of Tokyo, Tokyo 113-8656, Japan}

\author{Tan Van Vu}
\email{tan@biom.t.u-tokyo.ac.jp}
\affiliation{Department of Information and Communication Engineering, Graduate School of Information Science and Technology, The University of Tokyo, Tokyo 113-8656, Japan}

\author{Yoshihiko Hasegawa}
\email{hasegawa@biom.t.u-tokyo.ac.jp}
\affiliation{Department of Information and Communication Engineering, Graduate School of Information Science and Technology, The University of Tokyo, Tokyo 113-8656, Japan}
\allowdisplaybreaks

\maketitle
In this supplementary material, we show that the equality conditions of the derived TUR [Eq.~(24) in the main text] may be satisfied even in the far-from-equilibrium regime.
We consider a generalized Szilard engine with measurement errors \cite{Sagawa:2012:PRE}, which consists of the following five steps:
\begin{itemize}
\item \textit{Step 1}: Prepare a molecule in a box that has a volume equal to $1$.
    \item \textit{Step 2}: Insert a barrier into the middle of the box. We denote the location of the molecule after insertion by the state $x \in \{0,1\}$. 
    The state $x$ reads $0$ $(1)$ if the molecule is in the left (right) of the barrier.
    \item \textit{Step 3}: Measure the location of the molecule with an error probability $\epsilon \in (0,1)$. 
    The measurement outcome is denoted by $y\in \{0,1\}$. 
    The outcome $y$ reads $0$ $(1)$ if the molecule is measured to be in the left (right) of the barrier.
    \item \textit{Step 4}: According to the measurement outcome $y$, slowly expand the volume of the part in which the molecule is located until the volume reaches $v \in (1/2, 1)$.
    \item \textit{Step 5}: Remove the barrier from the box for the engine to return to its the initial state. 
\end{itemize}
The forward process is described by the joint probability distribution $P_{\rm F}(x,y)$ of the position and measurement outcome:
\begin{align}
    P_{\rm F}(x,y) \equiv \delta (x,y) (1- \epsilon) / 2+ (1 - \delta (x,y))  \epsilon /2,
\end{align}
where $\delta$ is the Kronecker delta ($\delta(x,y)=1$ for $x=y$, and $0$ otherwise).

To evaluate the extracted work, we consider the molecule as an ideal gas. In this case, the extracted work during this period can be found from $\beta^{-1}\int V^{-1}dV$, where $V$ and $\beta$ denote the volume and the inverse temperature, respectively. The extracted work is then given by
\begin{align}
	\beta  W(x,y)&= \ln(2) + \delta (x,y) \ln(v) + (1- \delta(x,y) )\ln(1 - v).
\end{align}
The mean and variance of the work can be calculated as
\begin{align}
	\beta \mean{W}_{\rm F} &=  \ln(2) + (1- \epsilon) \ln(v) +  \epsilon \ln (1 - v), \\
	\beta ^2 \var{W}_{\rm F} 
    &= \epsilon (1 - \epsilon ) \braket{\ln v - \ln(1-v)}^2.
\end{align}

The backward process is set up as in the following:
\begin{itemize}
    \item \textit{Step 1}: Prepare the box with a molecule.
    \item \textit{Step 2}: Insert the barrier into the box such that it divides the box into two parts with volumes $v$ and $1-v$.
    According to the measurement outcome $y = 0$ or $y = 1$ in the forward process, the left part has the volume $v$ or $1 - v$, respectively. 
    \item \textit{Step 3}: Slowly move the barrier toward the middle of the box.
    \item \textit{Step 4}: Measure the location of the molecule without error. The measurement outcome $x$ reads $0$ $(1)$ if the molecule is in the left (right) of the barrier.
    \item \textit{Step 5}: Remove the barrier from the box.
\end{itemize}
The joint probability distribution $P_B (x,y)$ of the position and measurement outcome in the backward process is given by
\begin{align}
    P_{\rm B}(x,y) &= \delta(x,y)v/2 + (1 - \delta(x,y))(1 - v)/2.
\end{align}
In the backward process, the extracted work for the given $x$, $y$ is equal to $-W(x,y)$. 
The mean and variance of the extracted work in the backward experiment are given by 
\begin{align}
   \beta \mean{W}_{\rm B} &=  -\ln(2) - v \ln(v) - (1- v) \ln(1 - v),\\
   \beta ^2\var{W}_{\rm B}  &= v(1-v) \braket{\ln v - \ln (1-v)}^2.
\end{align}
The irreversibility of the dynamics is evaluated using the probabilities of the forward and backward processes as
\begin{align}
	\sigma(x,y) \equiv \ln \frac{P_{\rm F}(x,y)}{P_{\rm B}(x,y)}. 
\end{align}
As noted in Ref.~\cite{Sagawa:2012:PRE}, the irreversibility can be decomposed as $\sigma(x,y) = -\beta W(x,y) + I[x:y] $, where $I[x:y] \equiv \ln \frac{P(y|x)}{P(y)}$ denotes the mutual information between $x$ and $y$.
The average of $\sigma$ over the forward distribution $P_{\rm F}(x,y)$ reads
\begin{align}
	\mean{\sigma}_{\rm F} &= D_{\rm KL}[P_{\rm F}(x,y)||P_{\rm B}(x,y) ] \\
	&=  \epsilon \ln \frac{\epsilon}{1-v} + (1-\epsilon) \ln \frac{1-\epsilon}{v}. \label{S_KL}
\end{align}

Applying the TUR derived in the main text for an observable of the extracted work, we find that $\mean{\sigma}_{\rm F}$ is lower bounded by 
\begin{align}\label{TUR}
  \int_0^1 \frac{\theta (\mean{W}_{\rm F} + \mean{W}_{\rm B})^2 }{\var{W}_\theta} d\theta &= \int_0^1 \frac{\theta (1- v - \epsilon)^2}{(1-\theta)\epsilon (1 - \epsilon ) + \theta v(1-v) +  \theta(1- \theta)(1- v - \epsilon)^2} d\theta\\
  	&= \int_0^1 -\frac{\epsilon(1 - \epsilon - v)}{\epsilon +(1- \epsilon - v) \theta} - \frac{(1-\epsilon)(1- \epsilon - v)}{(1-\epsilon) - (1- \epsilon - v) \theta} 	d\theta\\
	&= \epsilon \ln \frac{\epsilon}{1-v} + (1-\epsilon) \ln \frac{1-\epsilon}{v}. \label{S_TUR}
\end{align}
From Eqs.~\eqref{S_KL} and \eqref{S_TUR}, we obtain the following relation:
\begin{align}
	\mean{\sigma}_{\rm F} = \int_0^1 \frac{\theta (\mean{W}_{\rm F} + \mean{W}_{\rm B})^2 }{\var{W}_\theta} d\theta.
\end{align}
This indicates that the equality of our TUR can be attained even when the magnitude of $\mean{\sigma}_{\rm F}$ is arbitrarily large.